\documentclass[pra,aps,preprint,showpacs,12ppt]{revtex4}
\usepackage{graphicx,bm}

\newcommand{\grad}{\mbox{\boldmath$\nabla$}}

\begin{document}

\title{Strong-field approximation for harmonic generation in diatomic molecules}
\author{C.~C.~CHIRIL\u{A} and M.~LEIN}

\affiliation{Max Planck Institute for Nuclear Physics,
Saupfercheckweg 1, 69117 Heidelberg, Germany}

\date{\today}

\begin{abstract}
The generation of high-order harmonics in diatomic molecules is investigated
within the framework of the strong-field approximation. We show that the conventional saddle-point approximation is not suitable for large internuclear distances. An adapted saddle-point method that takes into account the molecular structure is presented. We analyze the predictions for the harmonic-generation spectra in both the velocity and the length gauge. At large internuclear separations, we compare the resulting cutoffs with the predictions of the simple-man's model. Good agreement is obtained only by using the adapted saddle-point method combined with the velocity gauge.
\end{abstract}


\maketitle

\section{Introduction}

The interaction of strong laser pulses with atoms or molecules leads to a variety of interesting phenomena. Among them, the generation of high-order harmonics \cite{McPherson,Huillier,Protopapas,Salieres} is of particular interest, as it serves as a source of high-frequency coherent radiation, along with the generation of attosecond pulses. The possibility of their control and optimization promises fascinating applications \cite{Kanai,Kienberger}.

  For atoms, harmonic generation (HG) has been intensively studied and is well understood in terms of the three-step semiclassical mechanism \cite{Corkum93,Kulander95}, also known as the simple-man's model. In the framework of the strong-field approximation (SFA), the simple-man's model has its quantum-mechanical counterpart in the model proposed by Lewenstein \emph{et al.}\ \cite{Lewenstein}. The Lewenstein model for HG in atoms provides a good qualitative and quantitative agreement with \emph{ab initio} calculations, which become prohibitive at high laser intensities. The advantage of the model is at least two-fold: First, in contrast to the integration of the time-dependent Schr\"odinger equation (TDSE), it provides a physical understanding of the underlying mechanisms and a certain degree of analytical description. Second, the HG spectrum can be obtained with much less computational effort than from the TDSE. This is essential for calculating macroscopic effects during the propagation of the harmonic radiation in the medium.

The study of HG in molecules is at its early stages. Since molecules have additional degrees of freedom and more complicated symmetries, the physical phenomena are much richer. Experiments have shown that the ionization of molecules is greatly influenced by the molecular orientation with respect to the laser polarization direction \cite{Litvinyuk03}. With the HG process depending greatly upon ionization, one has a similar effect for the emission of the harmonic radiation \cite{Kanai,Itatani,Velotta01}. Moreover, some molecules can withstand laser intensities much higher than in the atomic case \cite{Hankin00,Talebpour96&98}. As a consequence, harmonic radiation can be extended to higher photon energies \cite{Shan02}. Furthermore, in the semiclassical picture the returning wave packet encounters the molecular core with several atomic sites whose contributions interfere. Thus, the multicenter nature of molecules can lead to interference effects in HG \cite{Lein02,Lein02a}. Similarly, diffraction is predicted to occur in the above-threshold ionization (ATI) \cite{Zuo96,Lein02b} spectra. By varying the orientation of the molecule, a certain harmonic range can be maximized or minimized \cite{Lein02,Lein02a,Lein03,Kamta04,Kamta05}.

 A proposed mechanism to generate harmonics of very high frequency involves probing a molecule stretched well beyond its equilibrium distance \cite{Moreno}. Note that pump--probe experiments have already been carried out, where such large internuclear distances have been created \cite{Ergler05}. The large internuclear separation allows the electron to be detached from one site and accelerated by the laser field towards another site, where is captured and emits an energetic photon. According to the simple-man's model, the maximal kinetic energy before recombination can be as big as $8U_{\text p}$ in such a process \cite{Moreno,Kopold98}, where $U_{\text p}$ is the ponderomotive energy of the electron in the external field.

The theoretical description of HG in molecules is largely based up to now on the numerical integration of the Schr\"odinger equation in one spatial dimension \cite{Chelkowski96,Qu01,Castiglia04}, two dimensions \cite{Lappas00,Lein02a} or in three dimensions \cite{Zuo93,Lein03,Kamta04,Kamta05}. Other approaches include two-level models \cite{Zuo93,Plummer95}.  The numerical calculations for molecules require more care than for atoms since one needs to study the dependence on orientation and bond length. Including the vibrational motion of the nuclei poses yet another difficulty. Hence, developing theoretical models can be a way to alleviate these difficulties and provide deeper understanding of the physical mechanisms. The first choice would be the extension of the Lewenstein model to the molecular case. As it turns out, due to the multicenter aspect of molecules, this is not a straightforward task. For example, the ground state required in the Lewenstein model is not known analytically for molecules. One has to use approximations, such as the linear combination of atomic orbitals (LCAO) approximation. It was shown that the ground state must have the correct symmetry in order for the model to be able to describe at least at a qualitative level the dependence of ionization on the molecular orientation. This can be seen from the successful molecular-Ammosov-Delone-Krainov (MO-ADK) theory \cite{Tong02} for molecular ionization. Using the correct ground-state symmetry, the Lewenstein model for molecules seems to give good qualitative predictions \cite{Zhou05} for the alignment dependence of the HG yields. For a more detailed discussion on the importance of the orbital symmetry, see Ref.\ \cite{Madsen05}.

 The choice of gauge in analytical approximations is known to be difficult. While results from the TDSE are invariant under gauge transformations, this is not so for the SFA (for a discussion in the atomic case, see Ref.\ \cite{Bauer05}). The SFA was used in \cite{Kopold98} for a three-dimensional zero-range potential with two centers. Considering the harmonic emission from an atom which is displaced from the origin of the coordinate system by an arbitrary distance, the authors show that spurious effects, like the presence of even harmonics in the spectrum arise in both the velocity and the length gauge. While in  the velocity gauge the odd harmonics are \emph{invariant} under the spatial translation, in the length gauge this invariance holds for neither the even nor the odd harmonics. Moreover, for the calculation of the harmonic spectra there is a choice between dipole, momentum and acceleration formulations. In  Ref.\ \cite{Kamta05}, a simplified version of the Lewenstein model is adapted to the molecular case. The authors show that the two-center interference minima can be well reproduced provided that the acceleration formulation is used to calculate the HG spectrum (i.e., the dipole acceleration is calculated by means of the Ehrenfest's theorem, as the expectation value of the force acting on the active electron). The dipole formulation predicts interference minima in severe disagreement with the TDSE results.

In the present paper, we present a detailed analysis of the Lewenstein model for molecules, investigating both the length and the velocity gauge. We show that the conventional saddle-point approximation used for the Volkov evolution operator leads to unphysical results for large internuclear distances. We propose a modified version of the saddle-point approximation, taking into account the structure of the molecule. As a consequence, one obtains the HG amplitude as a sum over different electron trajectories, with ionization and recombination taking place at different molecular sites (the ``exchange term" from Ref.\ \cite{Kopold98}) or identical ones (``the return" term from Ref.\ \cite{Kopold98}). We show that only in the velocity gauge, the new saddle-point formulation gives agreement with the cutoffs predicted by the simple-man's model \cite{Moreno}.

 The paper is organized as follows. In Sec.\ \ref{theory} we revisit the conventional Lewenstein model and its extension to molecules, with the molecular ground state approximated as a linear combination of atomic orbitals. Both the length gauge and the velocity gauge are studied comparatively throughout this work. The resulting expressions are then analyzed within the conventional saddle-point approximation in Sec.\ \ref{SP}, and an adapted version of the saddle-point approximation is proposed. This new version takes into account the molecular structure, which now appears explicitly in the emerging electronic trajectories derived from the saddle-point equations. Sec.\ \ref{results} is devoted to the numerical evaluation of the different analytical expressions of the HG amplitude. The harmonic cutoffs for different internuclear distances are compared to the cutoffs predicted by the simple-man's model applied to molecules  \cite{Moreno}. It is shown, both analytically and numerically, that only the velocity gauge together with the adapted saddle-point approximation yields cutoffs in agreement with the simple-man's model. Finally, Sec.\ \ref{conclusions} contains a short summary, our conclusions and perspectives.

\section{Theoretical approach} \label{theory}

\subsection{The strong-field approximation (length gauge)}\label{LG}
To obtain the strong-field approximation for the harmonic-generation process, we follow closely the approach of the Lewenstein model of HG in atoms \cite{Lewenstein}. We start from the dipole approximation and the length--gauge Hamiltonian for a H$_2^{+}$ molecule with fixed orientation irradiated by a laser field linearly polarized along the $x$ axis (atomic units are used throughout):
\begin{equation}
  H = -{\grad^2\over2}+ V(\mathbf{r},\mathbf{R}) + E(t)x,
\end{equation}
where $V$ is the Coulomb interaction with the two protons and $\mathbf{R}$ is the internuclear distance, here taken as a parameter. Its orientation with respect to the laser polarization direction may vary. The molecule is in the $(x,y)$ plane and the laser propagates along the $z$ axis. We label the two nuclei as nucleus $A$ at the position $-\mathbf{R}/2$ and nucleus $B$ at position $\mathbf{R}/2$, respectively.

Following Ref.~\cite{Lewenstein}, we assume that (a) no bound states other
than the ground state are populated, 
(b) the depletion of the ground state
can be neglected, and (c) while in the continuum, the active
electron 
does not interact with the core. 
Although we focus on the hydrogen molecular ion, for molecules such as the hydrogen molecule, one may additionally assume 
that only a single electron can become active, i.e., if
one of the electrons has been excited into the continuum,
the second electron will not couple to the field and will
always remain in the lowest electronic state of the molecular ion. 
The expression for the electronic dipole moment is the same as given in \cite{Lewenstein} for an atom:
\begin{eqnarray}\label{dipole}
  \mathbf{D}(t) = -i\int\limits_{0\,}^{\,t}\!dt'\int\!\!{d^3\!p} \;
\mathbf{d}_{\rm{rec}}^{*}[\mathbf{p+A}(t)]d_{\rm{ion}}[\mathbf{p+A}(t'),t'] \, \exp[-iS(\mathbf p,t',t)]+\,{\rm c.c.},
\end{eqnarray}
with $S=\int_{t'}^tdt''\{[\mathbf{p}+\mathbf{A}(t'')]^2/2+I_{\text p}\}$ being the semiclassical action and $I_{\text p}$ the ionization potential of the electronic ground state. (The dependence on the internuclear distance $\mathbf{R}$ was dropped in Eq.\ (\ref{dipole}) for clarity. In the following we make this dependence explicit.) Here, $\mathbf{A}(t) = -\int_{-\infty}^t \mathbf{E}(t')dt'$. The HG spectrum of light polarized along a certain direction $\hat e$ is obtained by modulus squaring the Fourier transform of the dipole acceleration along that direction:
\begin{equation}\label{dipacc}
	\hat{e} \cdot \mathbf{a}(\Omega)=\int_0^{T_{\text p}}dt \, \hat{e} \cdot \ddot{\mathbf D}(t)\exp(i\Omega t),
\end{equation}
where the integration is carried out over the duration of the laser pulse, $T_{\text p}$. Due to the anisotropy of the molecular system, in contrast to atoms, the emitted radiation can be polarized along other directions than the laser polarization axis. Here we consider only the harmonic radiation polarized along the direction of the laser electric field, $\hat x$.

In Eq.\ (\ref{dipole}), $d_{\rm{ion}}$ is the ionization amplitude
\begin{equation}\label{dion}
d_{\rm{ion}}(\mathbf{k,R},t)=\langle\psi_V(\mathbf{k})\vert E(t)x \vert\psi_0(\mathbf{R})\rangle.
\end{equation}
It has the simple interpretation of the electron transition amplitude from the ground state to a Volkov state $\vert\psi_V(\mathbf{k})\rangle$ of momentum $\mathbf{k}$. For the calculation of $d_{\rm{ion}}$ in (\ref{dion}) and $\mathbf{d}_{\rm{rec}}$ below, we use only the spatial part of the Volkov solution. The temporal phase of the Volkov solution is included in the semiclassical action.

For the recombination step, we have
\begin{equation}\label{drec}
\mathbf{d}_{\rm{rec}}(\mathbf{k},\mathbf{R})=\langle\psi_V(\mathbf{k})\vert -\mathbf{r} \vert\psi_0(\mathbf{R})\rangle,
\end{equation}
whereby an electron under the influence of the external field only, described by a plane-wave Volkov solution, recombines with the molecular core. Because we employ plane waves to describe the electron motion in the continuum, some peculiarities in the predictions for the harmonic spectra are expected. The plane-wave description is more accurate for fast electrons returning to the ionic core after spending a longer time in the continuum. As a consequence, the predictions are expected to be  more accurate for high-order harmonics than for low-order ones, where other effects not included here, such as the influence of the binding potential, come into play. 
A shortcoming of the current SFA formulation is that the ionization and recombination amplitudes are not translationally invariant, due to the fact that the Volkov wave function and the ground state are not orthogonal. For atoms, the predictions of the SFA for harmonic generation are in good agreement with the \emph{ab initio} results, \emph{provided} that the atom is considered in the origin of the coordinate system. For molecules, the nuclei cannot be both in the origin of the coordinate system and the lack of translation invariance causes more serious problems. We discuss these in the later sections of this work.

In order to calculate the dipole moment, one needs to know the ground state of the system. For the H-atom, this is known analytically, but for molecules one has to use approximations to obtain an analytical form. Here we adopt the linear combination of atomic orbitals (LCAO) approximation, i.e., the molecular ground state is taken to be:
\begin{equation}\label{LCAO}
\psi_0(\mathbf{r,\mathbf{R}})=\frac{1}{\sqrt{2[1+s(R)]}}\left[\psi_h(\mathbf{r}+\mathbf{R}/2) + 
\psi_h(\mathbf{r}-\mathbf{R}/2) \right],
\end{equation}
with $\psi_h(\mathbf{r})$ being the hydrogen ground state. Furthermore, $s(R)=\exp(-R)(3+3R+R^2)/3$ is the overlap integral between the two atomic orbitals, needed to ensure normalization. By using Eq.\ (\ref{LCAO}) in (\ref{dion}) and (\ref{drec}), the ionization amplitude reads
\begin{equation}\label{dionLCAO}
d_{\rm{ion}}(\mathbf{k, R},t) = iE(t)\sqrt{ \frac{2}{1+s(R)} } \left[ -\frac{R_x}{2}\sin \bigg( \frac{\mathbf{k}\cdot\mathbf{R}}{2}\bigg)
\tilde{\psi_{h}}(\mathbf k)+\cos\bigg(\frac{\mathbf{k}\cdot\mathbf{R}}{2}\bigg) \frac{\partial\tilde\psi_{h}(\mathbf k)}{\partial k_x} \right],
\end{equation}
where $\tilde\psi_{h}(\mathbf k)$ is the Fourier transform of the hydrogen ground state wave function, $\tilde\psi_{h}(\mathbf k)=(\pi\sqrt{2})^{-1}(\mathbf{k}^2/2+1/2)^{-2}$. The recombination amplitude has a similar form
\begin{equation}\label{drecLCAO}
\mathbf{d}_{\rm{rec}}(\mathbf{k, R}) = -i\sqrt{ \frac{2}{1+s(R)} } \left[ -\frac{\mathbf{R}}{2}\sin \bigg( \frac{\mathbf{k}\cdot\mathbf{R}}{2}\bigg)
\tilde{\psi_{h}}(\mathbf k)+\cos\bigg(\frac{\mathbf{k}\cdot\mathbf{R}}{2}\bigg) \frac{\partial\tilde\psi_{h}(\mathbf k)}{\partial \mathbf{k}} \right].
\end{equation}
In comparison to the atomic case, the transition amplitudes (\ref{dionLCAO}) and (\ref{drecLCAO}) depend now on the internuclear distance $R$ and the molecular orientation. Thus, two-center interference effects \cite{Lein02} in the harmonic spectrum might arise.

Before proceeding to the next section, we briefly compare the exact bound state to the LCAO form that we use in the ionization and recombination amplitudes. The internuclear distance chosen here is the equilibrium distance in H$_2^+$.
\begin{figure}[!htb]
  \begin{center}
    \includegraphics[width=.6\textwidth,angle=-90]{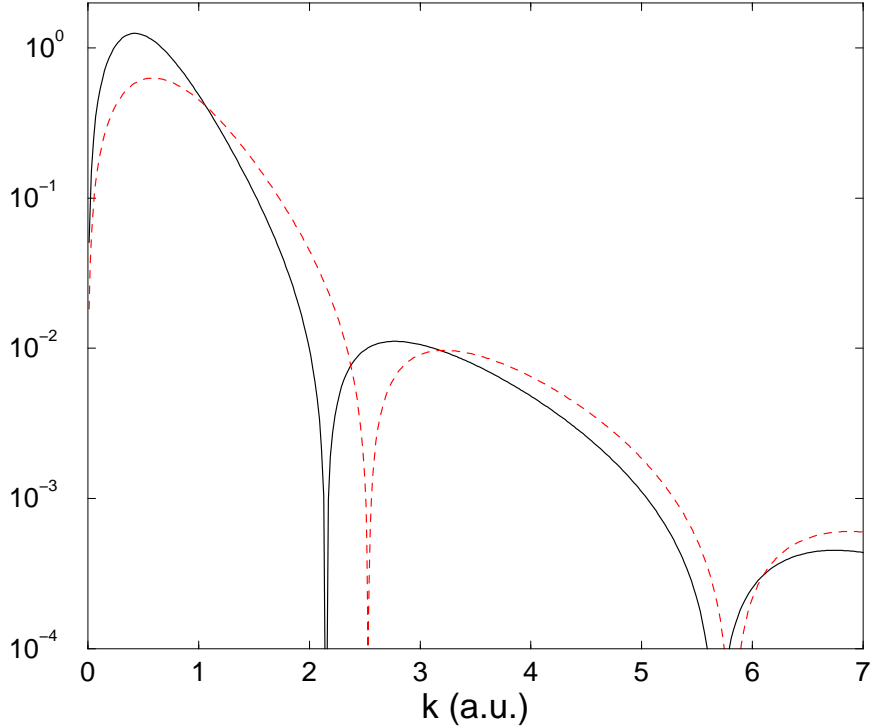}
  \end{center}
  \caption{The absolute value of the projection on the internuclear axis of  $\grad \tilde\psi_0(\mathbf k)$ is shown for the LCAO approximation (continuous curve) and the exact ground state (dashed curve) of H$_2^{+}$. The internuclear distance is $R=2$ atomic units.}
\label{fig:dpsi}
\end{figure}
Fig.\ \ref{fig:dpsi} shows the comparison between the gradient along the internuclear axis of the Fourier transform $\tilde\psi_{0}(\mathbf k)$ of the electronic ground state, obtained by using the LCAO wave function and, correspondingly, by using the `exact' one. The dips in the figure appear at zeros of the two functions. The projection of the gradient on the internuclear axis is useful when studying molecules aligned with the laser field. In this case the electron momentum is mainly directed along the molecular axis and the recombination amplitude relates to the quantity showed in the figure. One can see that the LCAO approximation works well.

\subsection{The strong-field approximation (velocity gauge)}\label{VG}

 To complete the analysis of the strong-field approximation, we adopt as well the velocity gauge. The Hamiltonian reads:
\begin{equation}
  H = {\big[-i\grad+\mathbf{A}(t)\big]^2\over2}+ V(\mathbf{r},\mathbf{R}),
\end{equation}
with $\mathbf{A}(t)$ the vector potential. Similar to the results in the length gauge (\ref{LG}), the dipole moment is:
\begin{eqnarray}\label{dipoleVG}
  \mathbf{D}(t) = -i\int\limits_{0\,}^{\,t}\!dt'\int\!\!{d^3\!p} \;
\mathbf{d}_{\rm{rec}}^{*}(\mathbf{p},t)d_{\rm{ion}}(\mathbf{p},t') \, \exp[-iS(\mathbf p,t',t)]+\,{\rm c.c.}
\end{eqnarray}
The differences with respect to the length-gauge form occur in the matrix elements describing the ionization and the recombination. This is due to the different form of the Volkov wave function and of the interaction Hamiltonian. For the ionization amplitude $d_{\rm{ion}}(\mathbf{k,R},t)=\langle\psi_V(\mathbf{k})\vert -i\grad\cdot \mathbf{A}(t) + \mathbf{A}^2(t)/2 \vert\psi_0(\mathbf{R})\rangle$ we obtain
\begin{equation}\label{dionLCAOvg}
d_{\rm{ion}}(\mathbf{k, R},t) = \sqrt{\frac{2}{1+s(R)}}\left[\mathbf{A}(t)\cdot\mathbf{k}+\frac{\mathbf{A}^2(t)}{2}\right]\cos\bigg(\frac{\mathbf{k}\cdot\mathbf{R}}{2}\bigg) \tilde\psi_{h}(\mathbf k).
\end{equation}
The recombination amplitude has the same functional form as in Eq.\ (\ref{drecLCAO}), but it appears in Eq.\ (\ref{dipoleVG}) with an argument different from Eq.\ (\ref{dipole}). The difference is that in the velocity gauge, the drift momentum $\mathbf p$ appears in the matrix elements, instead of the kinematical momentum $\mathbf{p+A}(t)$.

Although the length and the velocity gauge should be equivalent, significant differences appear in the predicted harmonic-generation spectra. Employing the saddle-point approximation, a detailed analysis of the two gauges is given in the following. We show that the presence of two centers makes the usual saddle-point approximation used to sum over the electron momenta in Eq.\ (\ref{dipoleVG}) questionable. This is due to the oscillatory terms present in the Fourier transform of the ground state [e.g., the factor $\cos(\mathbf{k}\cdot\mathbf{R}/2)$ in Eq.\ (\ref{dionLCAOvg})]. For large internuclear distances, it becomes necessary to take these terms into account when applying the saddle-point method.

\section{The saddle-point approximation and the corresponding electronic trajectories}\label{SP}

\subsection{Length gauge}

\subsubsection{The case of small internuclear distance}\label{LG small R}

 The dipole expression in (\ref{dipole}) contains an integral over all possible intermediate electron momenta $\mathbf p$. The presence of the semiclassical action $S(\mathbf p,t,t')$ makes the integrand highly oscillatory and therefore the saddle-point approximation can be used. \emph{Assuming} that the rest of the integrand is a \emph{smooth} function of momentum, the saddle-point momentum is obtained from the condition that the action is stationary, i.e., $\grad_{\mathbf p} S(\mathbf{p},t,t')\vert_{\mathbf{p=p}_s}=0$, implying $\mathbf{p}_s(t',t)=-\int_t^{t'}dt'' \mathbf{A}(t'')/(t-t')$. The dipole moment along the laser polarization direction becomes:
\begin{eqnarray}\label{dipoleSP}
  D_x(t) = -i\int\limits_{0\,}^{\,t}\!dt' \;\biggl[\frac{2\pi}{\epsilon+i(t-t')}\biggr]^{3/2}
d_{\rm{rec},x}^{*}[\mathbf{p}_s(t',t)+\mathbf{A}(t)] \, \exp\{-iS[\mathbf{p}_s(t',t),t',t]\} \nonumber\\
\times \, d_{\rm{ion}}[\mathbf{p}_s(t',t)+\mathbf{A}(t'),t']+\,{\rm c.c.},\hspace{5cm}
\end{eqnarray}
with $\epsilon$ a small cutoff parameter. The term $(t-t')^{-3/2}$ is related to the spreading of the electron wave packet during the motion in the continuum.

For a complete description in terms of complex electron trajectories \cite{Lewenstein95,Kopold00,Milosevic00,Milosevic00a} that contribute to the HG spectrum, one needs to apply the saddle-point approximation for both the integration over the time $t'$ in (\ref{dipoleSP}) and over $t$ in (\ref{dipacc}). This gives two more saddle equations, one for the \emph{tunnelling} time $t'_s$ and one for the harmonic \emph{emission} time $t_s$. As all the saddle equations are obtained from the semiclassical action, which is the same as in the atomic case, the electron trajectories do not reflect the structure of the molecule. The only differences lie in the expressions for the recombination and ionization amplitudes, but the saddle momentum which governs the trajectories in the simple-man's model \cite{Corkum93} does not `feel' the molecular structure. Although they are the same as in \cite{Lewenstein}, we repeat them here for completeness:
\begin{eqnarray}\label{lewenst}
 \frac{[\mathbf{p}_s(t'_s,t_s)+\mathbf{A}(t_s')]^2}{2}+I_{\text p}=0 \; , \nonumber\\
 \frac{[\mathbf{p}_s(t'_s,t_s)+\mathbf{A}(t_s)]^2}{2}=\Omega-I_{\text p}.
\end{eqnarray}
They simply state that the electron is born in the continuum with approximately  zero kinetic energy and that the energy is conserved when the electron recombines into the ground state, followed by emission of a photon with energy $\Omega$. Equations (\ref{lewenst}) can be used to determine the cutoff in the harmonic spectrum, which is predicted to occur for a monochromatic laser field close to $\Omega = 3.17 U_{\text p} + I_{\text p}$, the same as in the atomic simple-man's model.

Unlike the H-atom case, for molecules both the ionization (\ref{dionLCAO}) and the recombination (\ref{drecLCAO}) amplitude contain an oscillatory factor, whose argument is proportional to the internuclear distance $R$. This means that for increasing $R$, the integrand becomes more and more oscillatory as a function of $t$ and $t'$, i.e., the above saddle-point equations are not valid and the cutoff of the harmonic spectrum increases. We find this effect present in the numerical calculations. Moreover, the extra oscillatory factor makes the saddle-point method for momentum questionable. Hence, it becomes necessary to include the extra oscillations by adding them to the semiclassical action, when the saddle-point approximation is applied. We discuss this approach and its consequences in the following.

\subsubsection{The case of large internuclear distance}
\label{LG large R}

 To account for the extra oscillatory factors, we include them in the oscillatory function used to derive the saddle momentum. In the simpler method of the previous section, only the oscillations due to the semiclassical action were taken into account. The present procedure reveals a more accurate picture of the harmonic generation process. It changes the definition of the complex electron trajectories, which now depend on the structure of the molecule. We proceed by re-writing the ionization and recombination amplitudes in the following form:
\begin{eqnarray}\label{new saddle}
d_{\rm{ion}}(\mathbf{k,R},t)=i_{+}\exp(i\mathbf{k}\cdot\mathbf{R}/2)+
 i_{-}\exp(-i\mathbf{k}\cdot\mathbf{R}/2)\nonumber \\
d_{\rm{rec},x}(\mathbf{k,R})=r_{+}\exp(i\mathbf{k}\cdot\mathbf{R}/2)+
 r_{-}\exp(-i\mathbf{k}\cdot\mathbf{R}/2).
\end{eqnarray}
From the product $d^{*}_{\rm{rec},x}[\mathbf{p+A}(t),\mathbf{R}]d_{\rm{ion}}[\mathbf{p+A}(t'),\mathbf{R},t']$ of the two amplitudes (\ref{new saddle}) there are four terms arising for different combination of signs in the exponential. Each of these terms gives rise to a different oscillatory behavior of the integrand. Therefore, the saddle equation for the electron momentum is different for each of the four cases. We treat them separately  and emphasize the physical picture they convey.

\begin{enumerate}

\item 
The combination with the prefactor $r^{*}_{+} i_{-}$ leads to an oscillatory behavior through the phase $S(\mathbf{p},t',t) +\mathbf{p}\cdot \mathbf{R}+[\mathbf{A}(t)+\mathbf{A}(t')]\cdot \mathbf{R}/2$. The condition for the saddle momentum reads $\grad_{\mathbf p} S(\mathbf{p},t',t)\vert_{\mathbf{p=p}_s}=-\mathbf{R}$, i.e., the saddle momentum is $\mathbf{p}_s(t',t)=-\int_{t'}^t dt''\mathbf{A}(t'')/(t-t')-\mathbf{R}/(t-t')$. The absolute value of the integral over $A(t'')$ can be estimated to be smaller than $2\alpha_0$, where $\alpha_0=E_0/\omega^2$ is the classical amplitude of oscillation of the electron in the external field. Thus it becomes apparent that the present modification of the saddle-point method is certainly required for $R>2\alpha_0$, giving us a more precise definition of \emph{large} internuclear distances.

Applying the saddle condition to perform the integration over $t'$ in Eq.\ (\ref{dipole}) and over $t$ in Eq.\ (\ref{dipacc}), one obtains for the \emph{tunnelling} time $t'_s$ of the electron in the continuum
\begin{equation}\label{t'sBA}
\frac{[\mathbf{p}_s(t'_s,t)+\mathbf{A}(t')]^2}{2}+I_{\text p}+\mathbf{E}(t'_s)\cdot\mathbf{R}/2=0,
\end{equation}
and for the \emph{emission} time $t_s$ of a harmonic with frequency $\Omega$ 
\begin{equation}\label{tsBA}
\frac{[\mathbf{p}_s(t'_s,t_s)+\mathbf{A}(t_s)]^2}{2}=\Omega-I_{\text p}+\mathbf{E}(t_s)\cdot\mathbf{R}/2.
\end{equation}
The expression of the saddle momentum along with the condition for the tunnelling time (\ref{t'sBA}) define an electron trajectory such that, after the electron was `born' in the continuum at nucleus $B$ at a time close to Re$(t'_s)$, it oscillates in the field and recombines with the molecular core at nucleus $A$. We call the harmonics generated by tunneling and recombination at different molecular sites \emph{transfer} harmonics.

Equations (\ref{t'sBA}) and (\ref{tsBA}) are conditions for the conservation of energy during ionization and recombination, respectively. In these equations, the ionization potential is unchanged (remember that the SFA considers the ground state not dressed by the external field). Only the continuum is shifted by the potential energy of the electron in the external electric field at the two sites, A and B. This way, there is an additional unphysical energy difference between the Volkov state and the ground state of the electron, which can be seen in Eqns.\ (\ref{t'sBA}) and (\ref{tsBA}).

Similarly, the term with the prefactor $r_{-}^{*}i_{+}$ describes a process where the electron is `born' at the nucleus $A$ and recombines at the nucleus $B$. The equations describing this process are as given above, with $\mathbf{R}$ changing sign.

\item 
The combination with the prefactor $r^{*}_{+} i_{+}$ leads to the phase $S(\mathbf{p},t',t)+[\mathbf{A}(t)-\mathbf{A}(t')]\cdot \mathbf{R}/2$. The condition for the saddle momentum reads $\grad_{\mathbf p} S(\mathbf{p},t',t)\vert_{\mathbf{p=p}_s}=0$, i.e., the saddle momentum is $\mathbf{p}_s(t',t)=-\int_{t'}^t dt''\mathbf{A}(t'')/(t-t')$. For the \emph{tunnelling} time one obtains:
\begin{equation}\label{t'sBA1}
\frac{[\mathbf{p}_s(t'_s,t)+\mathbf{A}(t')]^2}{2}+I_{\text p}-\mathbf{E}(t'_s)\cdot\mathbf{R}/2=0.
\end{equation}
The expression of the saddle momentum along with (\ref{t'sBA1}) define an electron trajectory such that, after the electron was `born' in the continuum at nucleus $A$ at a time close to Re$(t'_s)$, it oscillates in the field and recombines with the molecular core at the same nucleus (we call these \emph{direct} harmonics). The saddle condition for the emission time $t_s$ reads 
\begin{equation}\label{tsBA1}
\frac{[\mathbf{p}_s(t'_s,t_s)+\mathbf{A}(t_s)]^2}{2}=\Omega-I_{\text p}+\mathbf{E}(t_s)\cdot\mathbf{R}/2.
\end{equation}
 For the ionization and recombination of the electron at the nucleus $A$, the ionization potential is unaffected by the field, but the continuum is shifted by the potential energy $\mathbf{E}\cdot\mathbf{R}/2$ of the electron in the external electric field. 

The term with the prefactor $r_{-}^{*}i_{-}$ corresponds to an electron `born' at nucleus $B$ and recombining with the same nucleus. The relations describing this process are as given above, with $\mathbf{R}$ changing sign.
\end{enumerate}

The saddle equations for the emission time, Eqns.\ (\ref{tsBA}) and (\ref{tsBA1}), state that the relation between the energy of the returning electron and the harmonic frequency of the emitted radiation depends on the continuum energy shifts $\mathbf{E}(t'_s)\cdot \mathbf{R}/2$ and $\mathbf{E}(t_s)\cdot \mathbf{R}/2$. For large values of $\bm{R}$, this gives rise to unphysically high harmonic cutoffs. One can make a rough estimate of the cutoff for large R, obtaining $\Omega_{\rm{c}}$ of the order $\mathcal{O}(E_0 R)$, where $E_0$ is the electric field amplitude. 

In conclusion, the harmonic cutoff still increases monotonically with increasing internuclear distance. In the velocity gauge, the new formulation of the saddle-point approximation for momenta eliminates this artifact.

\subsection{Velocity gauge}

\subsubsection{The case of small internuclear distance}

 Using the velocity-gauge expressions for the ionization and recombination amplitudes from Sec.\ \ref{VG}, and employing the conventional saddle-point approximation for the electron momenta, one obtains the dipole moment along the laser polarization direction
\begin{eqnarray}\label{dipoleSPVG}
  D_x(t) = -i\int\limits_{0\,}^{\,t}\!dt' \;\biggl[\frac{2\pi}{\epsilon + i(t-t')}\biggr]^{3/2}
d_{\rm{rec,x}}^{*}[\mathbf{p}_s(t',t)] \, \exp \{-iS[\mathbf{p}_s(t',t),t',t)]\} \nonumber\\
\times \, d_{\rm{ion}}[\mathbf{p}_s(t',t),t']+\,{\rm c.c.},\hspace{5cm}
\end{eqnarray}
with $\epsilon$ a small cutoff parameter. The saddle momentum is the same as in the length gauge. Again, the ionization and recombination amplitudes contain oscillatory factors that give rise to harmonic cutoffs that increase with increasing internuclear distance. The situation is completely different if we take the extra oscillatory factors into account when we apply the saddle-point approximation.

\subsubsection{The case of large internuclear distance}\label{VG large R}
 We proceed similarly to the length-gauge calculation,  re-writing the ionization and recombination amplitudes in the form of Eqns.\ (\ref{new saddle}). From the product $d^{*}_{\rm{rec},x}(\mathbf{k,R})d_{\rm{ion}}(\mathbf{k,R},t)$ of the two amplitudes  there are four terms arising for different combination of signs in the exponential. Each of these terms gives rise to a different oscillatory behavior of the integrand. Therefore, the saddle equation for the momentum of the electron is different for each of the four cases. We treat them separately and analyze the corresponding physical picture in terms of electronic trajectories.

\begin{enumerate}

\item 
The combination with the prefactor $r^{*}_{+} i_{-}$ leads to the phase $S(\mathbf{p},t',t) +\mathbf{p}\cdot \mathbf{R}$. The condition for the saddle momentum reads $\grad_{\mathbf p} S(\mathbf{p},t',t)\vert_{\mathbf{p=p}_s}=-\mathbf{R}$, i.e., the saddle momentum is $\mathbf{p}_s(t',t)=-\int_{t'}^t dt''\mathbf{A}(t'')/(t-t')-\mathbf{R}/(t-t')$. Using saddle-point approximations to perform the integration over $t'$ in Eq.\ (\ref{dipole}) and over $t$ in Eq.\ (\ref{dipacc}), one obtains for the \emph{tunnelling} time $t'_s$ of the electron
\begin{equation}\label{t'sBAVG}
\frac{[\mathbf{p}_s(t'_s,t_s)+\mathbf{A}(t'_s)]^2}{2}+I_{\text p}=0,
\end{equation}
and for the \emph{emission} time $t_s$ of a harmonic with frequency $\Omega$
\begin{equation}\label{tsBAVG}
\frac{[\mathbf{p}_s(t'_s,t_s)+\mathbf{A}(t_s)]^2}{2}=\Omega-I_{\text p}.
\end{equation}
The expression of the saddle momentum along with the condition for the tunnelling time (\ref{t'sBAVG}) define an electron trajectory such that, after the electron was `born' in the continuum at nucleus $B$ at a time close to $Re(t'_s)$, it oscillates in the field and recombines with the molecular core at nucleus $A$.
 
Similarly, the term with the prefactor $r_{-}^{*}i_{+}$ describes a process where the electron is `born' at nucleus $A$ and recombines at nucleus $B$. The equations describing this process are as given above, with $\mathbf{R}$ changing sign. We note that unlike for the case of the length gauge, the saddle equations that describe the complex trajectories agree with the classical equations of motions of the simple-man's model if one sets $I_{\text p}=0$, as is usually done when casting the complex trajectories in the language of the simple-man's model. The harmonic cutoff in this case can be as much as $8 U_{\text p} + I_{\text p}$ for a monochromatic field \cite{Moreno,Kopold98}. This will be addressed in more detail in the next section.

\item 
The combination with the prefactor $r^{*}_{+} i_{+}$ leaves the semiclassical action unchanged. The condition for the saddle momentum reads $\grad_{\mathbf p} S(\mathbf{p},t',t)\vert_{\mathbf{p=p}_s}=0$, i.e., the saddle momentum is $\mathbf{p}_s(t',t)=-\int_{t'}^t dt''\mathbf{A}(t'')/(t-t')$, like in the atomic case. For the \emph{tunnelling} time one obtains:
\begin{equation}\label{t'sBA1VG}
\frac{[\mathbf{p}_s(t'_s,t_s)+\mathbf{A}(t'_s)]^2}{2}+I_{\text p}=0.
\end{equation}
The expression of the saddle momentum along with (\ref{t'sBA1VG}) define an electron trajectory such that, after the electron was `born' in the continuum at a certain nucleus at a time close to $Re(t'_s)$, it oscillates in the field and recombines with the molecular core at the same nucleus. The saddle condition for the harmonic emission time $t_s$ reads 
\begin{equation}\label{tsBA1VG}
\frac{[\mathbf{p}_s(t'_s,t_s)+\mathbf{A}(t_s)]^2}{2}=\Omega-I_{\text p}.
\end{equation}
The term with the prefactor $r_{-}^{*}i_{-}$ corresponds to the same kind of trajectory, but for the other nucleus. The saddle equations are identical to the ones that describe the harmonic generation in the atomic case, and therefore have the usual cutoff of $3.17 U_{\text p} + I_{\text p}$ for stationary laser fields.
\end{enumerate}

In the velocity gauge, the new formulation of the saddle-point for momenta \emph{eliminates} the artifact that the harmonic cutoff increases indefinitely with increasing internuclear distance.

\section{Results and discussion}\label{results}

For the calculation of harmonic spectra, we choose a laser electric field with two optical cycles turn-on and turn-off and a plateau of three constant-amplitude optical cycles. Unless stated otherwise, we consider a H$_2^{+}$ molecular ion aligned with the field, irradiated by a laser with a wavelength of $800$ nm and intensity equal to $2 \times 10^{14}$ W/cm$^{2}$. The energy of the electronic ground state is given by $I_{\text p}(R) = V_{\rm{BO}}^{+}(R)-1/R$, where $V_{\rm{BO}}^{+}$ is the Born-Oppenheimer potential for the electronic ground state (including the proton-proton interaction energy).

\begin{figure}[!h]
  \begin{center}
    \includegraphics[width=.6\textwidth,angle=-90]{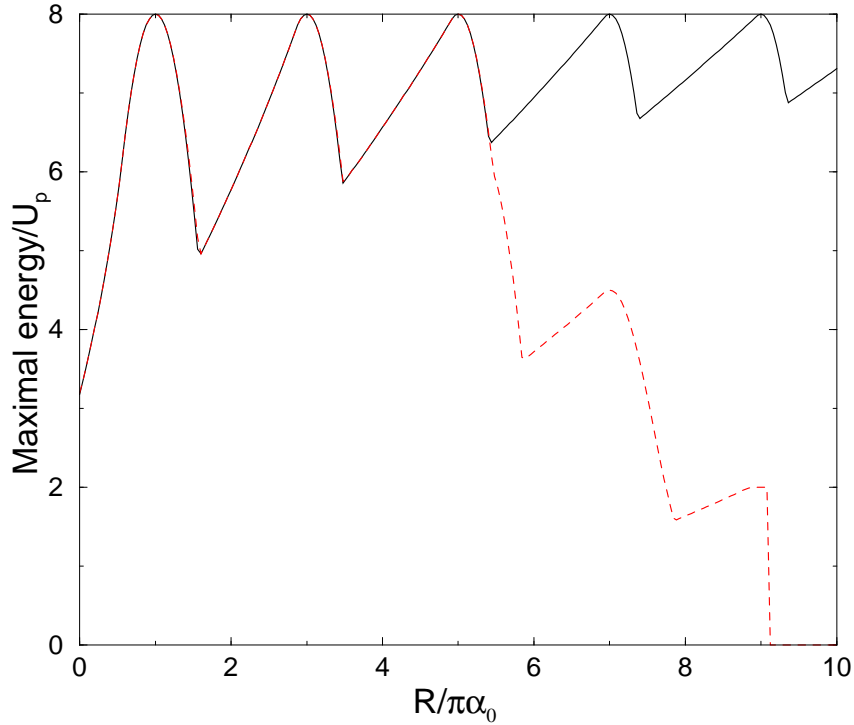}
  \end{center}
  \caption{The maximal kinetic energy upon arriving at nucleus B of an electron detached with zero initial kinetic energy from the nucleus A, as a function of the internuclear distance. Full curve, result for a monochromatic laser field, see Ref.\ \cite{Moreno}. Dashed curve, result for a trapezoidal envelope (see text).}
\label{fig:hgcutoffs}
\end{figure}

We focus our analysis on the variation of the cutoff energy with the internuclear distance $R$ \cite{Moreno,Kopold98}. In particular, we consider the cutoff related to the process in which the electron leaves the nucleus A and recombines at the nucleus B. As discussed in \cite{Kopold98}, for $R=(2n+1)\pi\alpha_0$ ($n=0,1,2\dots$) the maximal kinetic energy at B can be as much as $8 U_{\text p}$ for a linearly polarized monochromatic field, where $\alpha_0$ is large on the atomic scale (for the present laser parameters, $\alpha_0 = 23.3$ a.u.). For the maximal kinetic energy, the value of the internuclear distance is such that an electron detached at zero electric field at A is accelerated for half an optical cycle before reaching B. This way, the energy at recombination is much larger than $3.17 U_{\text p}$, as shown in Fig.\ \ref{fig:hgcutoffs}.

In the following, we analyze the harmonic spectra for both the length gauge and the velocity gauge, using the conventional and the adapted saddle-point method for integration over the electron momenta. 

\subsection{Length gauge results}

\begin{figure}[!h]
  \begin{center}
    \includegraphics[width=.6\textwidth,angle=-90]{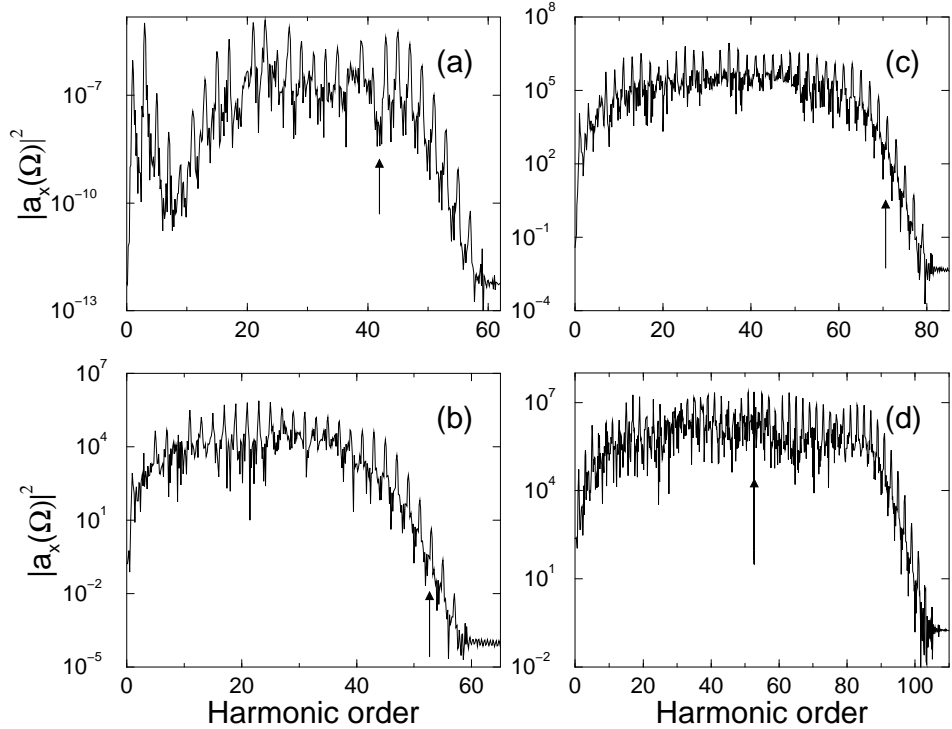}
  \end{center}
  \caption{Calculated harmonic spectra using the length gauge and the conventional saddle-point  method (see text), for 
various internuclear distances in H$_2^{+}$: (a) $R=0$ (the atomic case, $I_{\text p}=1$); (b) $R=0.5 \, \pi\alpha_0$;
 (c) $R=\pi\alpha_0$ and (d) $R=1.5 \, \pi\alpha_0$. The laser and molecule parameters are described in the beginning of Sec.\ \ref{results} and the arrows show the cutoffs of the transfer harmonics calculated from the simple-man's model.}
\label{fig:LG1}
\end{figure}

To better understand the harmonic spectra as they emerge from the SFA, panel (a) of Fig.\ \ref{fig:LG1} shows the harmonic spectrum for the atomic case. Using the conventional saddle-point method, the contribution of the recombination amplitude to the dipole moment simplifies to:
\begin{equation}
\label{eq: taylor}
a_x(\Omega) \propto (\Omega-I_{\text p}+1/2)^{-2} \left[R_x\sin\left(\frac{R_x k_{\rm{ret}}}{2}\right) -2\cos\left(\frac{R_x k_{ret}}{2}\right)\frac{k_{\rm{ret}}}{\Omega-I_{\text p}+1/2}\right],
\end{equation}
where $R_x$ is the projection of the internuclear distance on the polarization direction and $k_{\rm{ret}}$ is the momentum of the returning electron, $k_{\rm{ret}}=\sqrt{2(\Omega-I_{\text p})}$. From Eq.\ (\ref{eq: taylor}), it is seen that the behavior of the harmonic efficiency for harmonic orders \emph{bigger} than the molecular binding potential is oscillatory, while for \emph{smaller} harmonics is exponential. Thus, the predictions of the SFA in the energy region around the binding potential $I_{\text p}$ are distorted. This energy region is not described well by the SFA model, which does not fully include the effect of the Coulomb potential. Another remark in view of the above is that the two-center interference effect in HG is seen to stem mainly from the explicit form of the recombination amplitude. Therefore, for a good description of the interference, one has to provide a realistic expression for the recombination amplitude.

Fig.\ \ref{fig:LG1} shows harmonic spectra for different internuclear distances. The first thing to notice is that both the harmonic cutoff and the harmonic intensity increase with increasing internuclear separation. 
This is due to the presence of the oscillatory factors depending on $R$ in the ionization and recombination amplitudes (see the comments at the end of Sec.\ \ref{LG small R}). The arrows in Fig.\ \ref{fig:LG1} show the cutoff of the transfer harmonics as predicted by the simple-man's model (Fig.\ \ref{fig:hgcutoffs}). Since the present SFA formulation does not include rescattering of the electron from the molecular sites, we expect that the maximal harmonic order comes from the transfer mechanism. As seen from the figure, for large values of $R$, there is a significant difference between the SFA predictions in in the length gauge and the simple-man's predictions. The following figures will again display the cutoffs from the simple-man's model for comparison.

The contribution from the ionization amplitude in the case when $R$ is small depends on terms such as $\sinh (R_x \sqrt{2 I_{\text p}}/2)$ and $\cosh(R_x \sqrt{2 I_{\text p}}/2)$, as it can be shown by expanding the relevant quantities in Taylor series around the ionization times from the simple-man's model. This explains the increase of the harmonic intensity with increasing internuclear distance. Thus, the unphysical increase in the harmonic efficiency is exponential in $R$.

\begin{figure}[!h]
  \begin{center}
    \includegraphics[width=.6\textwidth,angle=-90]{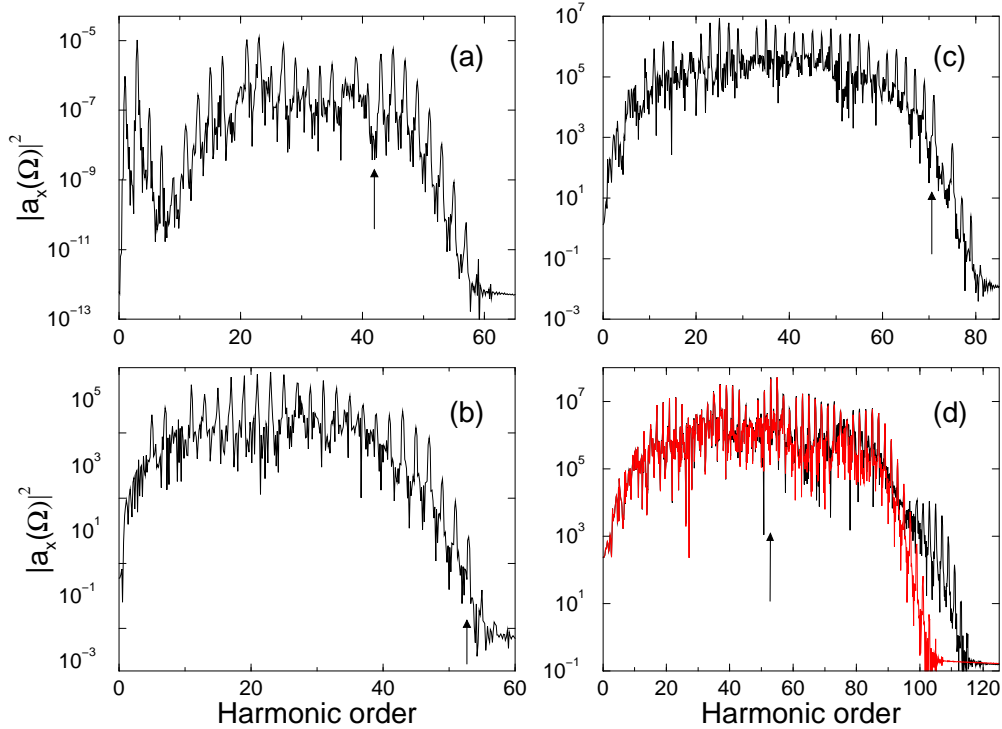}
  \end{center}
  \caption{ Harmonic spectra (same parameters as in Fig.\ \ref{fig:LG1}), but calculated by using the adapted saddle-point method and the length gauge. The lower curve in panel (d) shows the direct harmonics only. The arrows show the cutoffs of the transfer harmonics calculated from the simple-man's model (see Fig.~\ref{fig:hgcutoffs}).}
\label{fig:LG2}
\end{figure}

So far, we have discussed the results within the conventional saddle-point method for summation over the electron momenta. We turn now to the second saddle-point formulation for electron momenta (presented in Sec.\ \ref{LG large R}). Fig.\ \ref{fig:LG2} plots the predicted harmonic spectra.
To assess the importance of the direct harmonics, we plot them in panel (d) of Fig.\ \ref{fig:LG2}. Their cutoff is smaller in this case than the cutoff of the transfer harmonics. The artifacts regarding the cutoff and the harmonic efficiency are present.

\subsection{Velocity gauge results}

\begin{figure}[!h]
  \begin{center}
    \includegraphics[width=.6\textwidth,angle=-90]{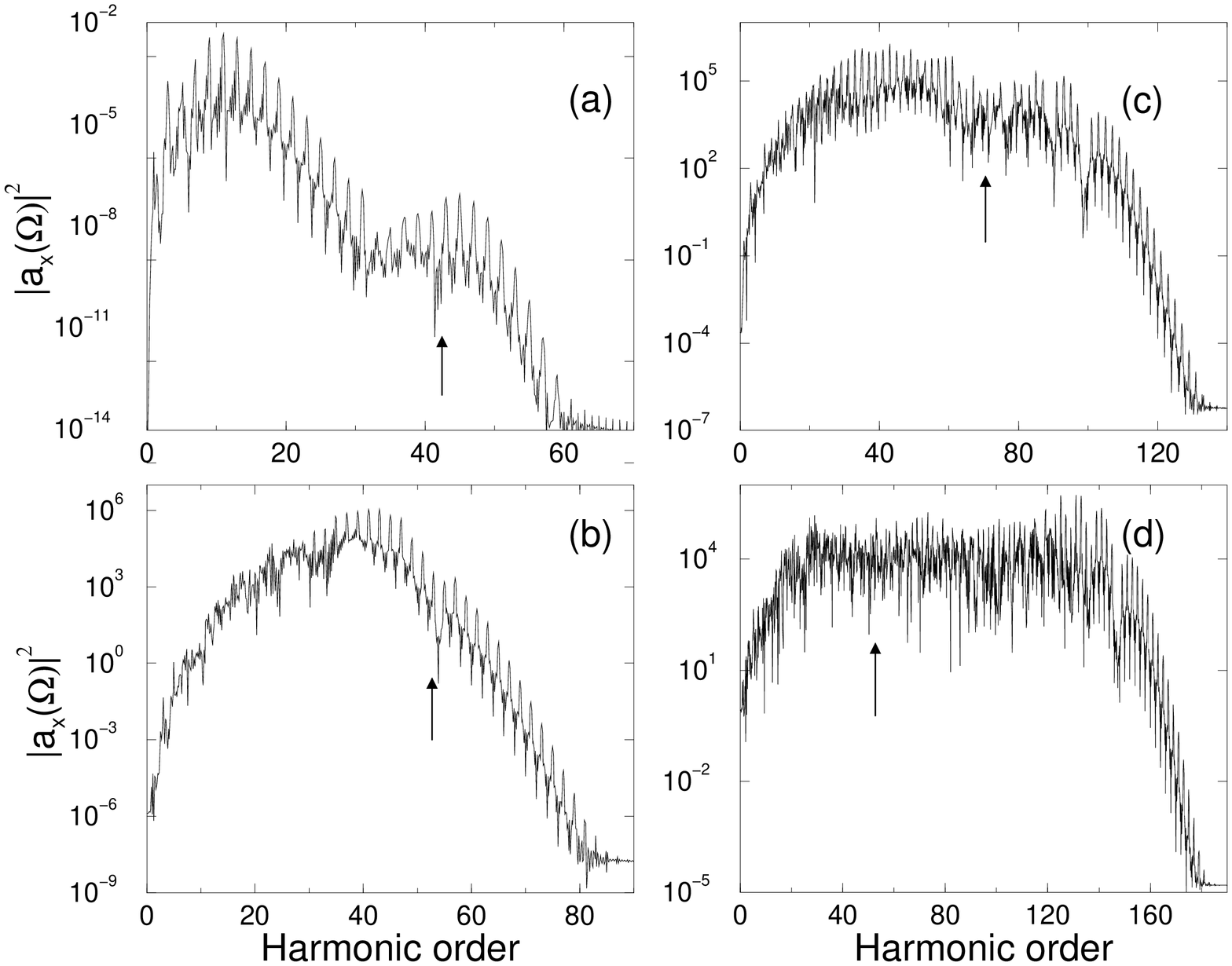}
  \end{center}
  \caption{Calculated harmonic spectra using the velocity gauge and the conventional saddle-point  method (see text), for 
various internuclear distances in H$_2^{+}$: (a) $R=0$ (the atomic case, $I_{\text p}=1$); (b) $R=0.5 \, \pi\alpha_0$;
 (c) $R=\pi\alpha_0$ and (d) $R=1.5 \, \pi\alpha_0$. The laser and molecule parameters are described in the beginning of Sec.\ \ref{results}. The arrows show the cutoffs of the transfer harmonics calculated from the simple-man's model.}
\label{fig:VG}
\end{figure}

\begin{figure}[!h]
  \begin{center}
    \includegraphics[width=.6\textwidth,angle=-90]{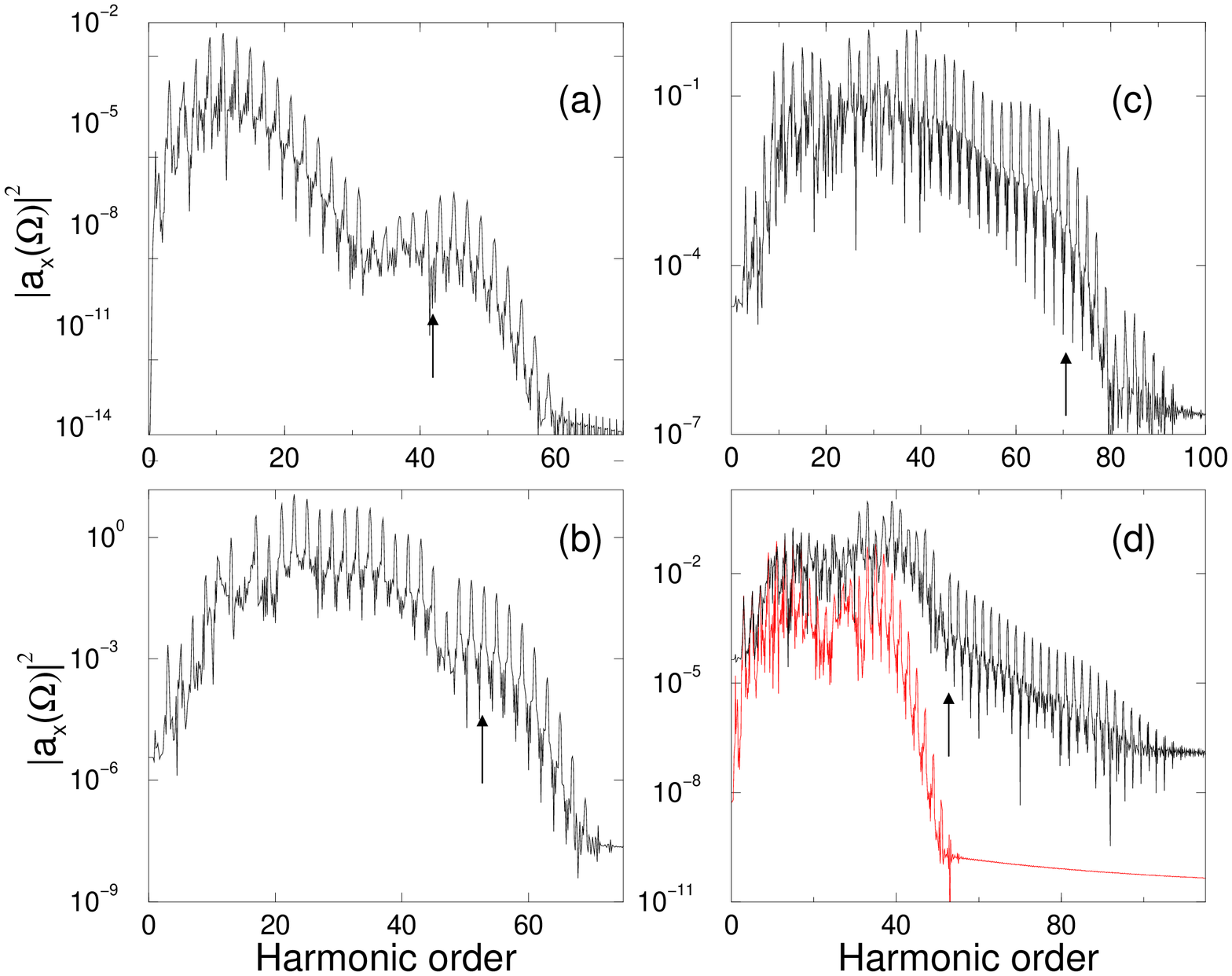}
  \end{center}
  \caption{ Harmonic spectra (same parameters as in Fig.\ \ref{fig:LG1}), but calculated by using the adapted saddle-point method and the velocity gauge. The lower curve in panel (d) shows the direct harmonics only. The arrows show the cutoffs of the transfer harmonics calculated from the simple-man's model.}
\label{fig:VG2}
\end{figure}

 For the velocity gauge and using the conventional saddle-point method, the harmonic spectra (Fig.\ \ref{fig:VG}) are similarly unphysical as those obtained in the length gauge. The shape of the spectra is slightly changed together with the values of the cutoffs. The increase of the cutoff with increasing internuclear distance $R$ is present.
A striking difference with respect to the length gauge, however, arises when one uses the second formulation of the saddle-point method, see Fig.\ (\ref{fig:VG2}). In this case, the unphysical increase of the cutoff with $R$ is eliminated.

\subsection{Comparison of the harmonic generation cutoffs with the predictions of the simple-man's model for large internuclear distances}

\begin{figure}[!h]
  \begin{center}
    \includegraphics[width=.6\textwidth,angle=-90]{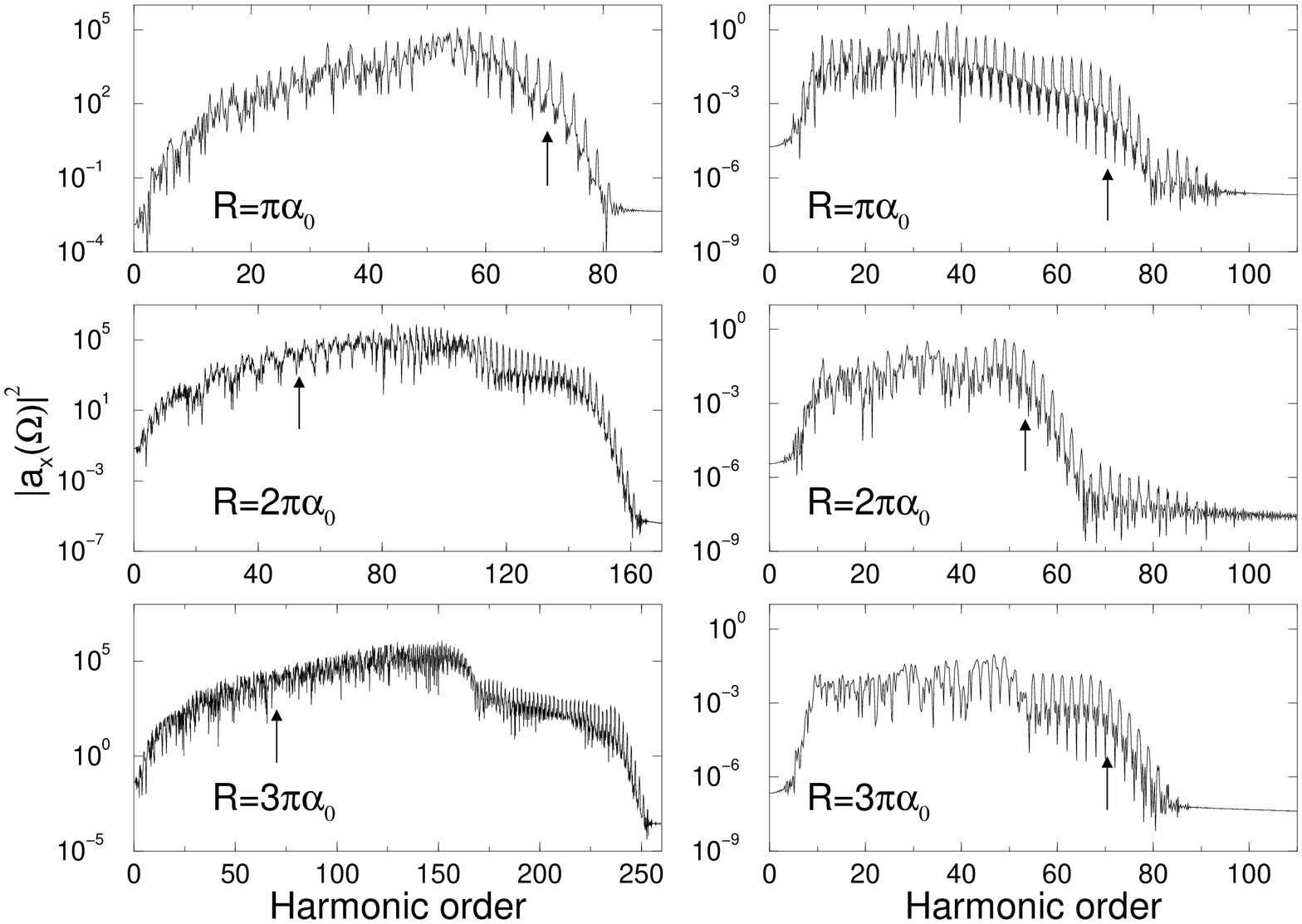}
  \end{center}
  \caption{ Transfer harmonics (same parameters as in Fig.\ \ref{fig:LG1}), calculated using the adapted saddle-point method in the length gauge (left column) and the velocity gauge (right column). The arrows show the cutoffs of the transfer-harmonics spectrum, predicted by the simple-man's model.}
\label{fig:SFAcutoffs}
\end{figure}

We compare the cutoffs for the transfer harmonics as predicted by the SFA using the adapted saddle-point method in both the length and velocity gauge with the predictions of the simple-man's model. Fig.\ \ref{fig:SFAcutoffs} shows the numerical results for very large internuclear separations. The arrows point to the semiclassical cutoffs from the simple-man's model. One can see that the length gauge overestimates by far the cutoffs as well as the harmonic intensity. In contrast, the velocity gauge together with the adapted saddle-point formulation succeeds in reproducing the simple-man's cutoffs, as well as keeping a bounded magnitude for the harmonic intensities.

\section{Conclusions and perspectives}\label{conclusions}

The SFA is a useful tool for studying processes in intense laser fields. It is of big advantage for molecules, where a solution of the Schr\"odinger equation can be more prohibitive than for an atom. There are still open questions about how to adapt the SFA for molecules such that its level of prediction would become as good as for the atomic case.

Here, we have analyzed the strong-field approximation for the calculation of the harmonic-generation spectra in ultrashort laser pulses. Using the conventional saddle-point method to approximate the sum over the electron momenta in the Volkov propagator, we have shown that both the length gauge and the velocity gauge predict harmonic intensities and cutoffs that increase unphysically with increasing internuclear separation. Also, the emerging trajectories for the electronic complex orbits \cite{Lewenstein95,Kopold00,Milosevic00,Milosevic00a} do not reflect the structure of the molecule and are the same as for the atomic case. In order to account for the presence of the two centers, we have introduced an adapted saddle-point method that takes into account the molecular structure. As a consequence, we obtain explicitly different types of electronic trajectories and corresponding harmonics, namely \emph{direct} harmonics (generated by tunneling and recombination of the active electron at the same site) and \emph{transfer} harmonics (generated by tunneling followed by recombination at a different site). The interesting observation is that the \emph{length gauge} still predicts in the adapted saddle-point formulation unbounded cutoffs and harmonic intensities as the internuclear separation is increased. Thus, the length-gauge SFA, in its present form, appears to be inappropriate for the description of high-harmonic generation at large internuclear distances. In contrast, the \emph{velocity gauge} predicts bounded harmonic intensities and cutoffs of the transfer harmonics. Only in the velocity gauge the cutoffs agree with the values obtained from the simple-man's model.

 Not addressed in this work is the question about the role played by the excited molecular electronic states in the harmonic emission process and whether or not it is necessary to include them in the SFA. Furthermore, we have not investigated the question which gauge is suitable for small internuclear separations, but there are indications that the length gauge will be preferable in this case \cite{Madsen04}. We believe that this work is a step towards building an accurate SFA for harmonic generation in molecules.

\end{document}